# A multichannel front end ASIC for PMT readout in LHAASO WCDA

**Yu Liang,**[a,b] **Lei Zhao,**[a,b,*] **Yuxiang Guo,**[a,b] **Jiajun Qin,**[a,b] **Yunfan Yang,**[a,b] **Boyu Cheng,**[a,b] **Shubin Liu,**[a,b] **Qi An**[a,b]

[a] *State Key Laboratory of Particle Detection and Electronics, University of Science and Technology of China, Hefei 230026, China*

[b] *Department of Modern Physics, University of Science and Technology of China, Hefei 230026, China*

   *E-mail:* zlei@ustc.edu.cn

ABSTRACT: Time and charge measurements over a large dynamic range from 1 Photo Electron (P.E.) to 4000 P.E. are required for the Water Cherenkov Detector Array (WCDA), which is one of the key components in the Large High Altitude Air Shower Observatory (LHAASO). To simplify the circuit structure of the readout electronics, a front end ASIC was designed. Based on the charge-to-time conversion method, the output pulse width of the ASIC corresponds to the input signal charge information while time information of the input signal is picked off through a discriminator, and thus the time and charge information can be digitized simultaneously using this ASIC and a following Time-to-Digital Converter (TDC). To address the challenge of mismatch among the channels observed in the previous prototype version, this work presents approaches for analyzing the problem and optimizing the circuits. A new version of the ASIC was designed and fabricated in the GLOBALFOUNDRIES 0.35 μm CMOS technology, which integrates 6 channels (corresponding to the readout of the 3 PMTs) in each chip. The test results indicate that the mismatch between the channels is significantly reduced to less than 20% using the proposed approach. The time measurement resolution better than 300 ps is achieved, and the charge measurement resolution is better than 10% at 1 P.E., and 1% at 4000 P.E., which meets the application requirements.

---

* Corresponding author.

# Contents

## 1. Introduction

The Large High Altitude Air Shower Observatory (LHAASO) is a multipurpose project, which is proposed to find the origin of cosmic ray and the very high energy gamma ray sources [1]. One of the key detectors in LHAASO is the Water Cherenkov Detector Array (WCDA) [2], which is designed to survey gamma ray sources at energies greater than 300 GeV. The total area of the WCDA is around 80,000 $m^2$, and consists of three water ponds. A total of 3120 Photomultiplier Tubes (PMTs) are placed under the water of these ponds to detect the Cerenkov light emitted by the extensive air shower (EAS) in the water. Every 9 adjacent PMTs are read out by one Front-End Electronics module (FEE). Each FEE can perform both high precision time and charge measurements over a large dynamic range which can lie between 1 Photo Electron (P.E.) and 4000 P.E [3].

In order to meet such measurement requirements, a prototype of the front end ASIC [4] which is based on the Time-Over-Threshold (TOT) method [5] was designed. All the analog front end circuits are integrated in this ASIC and its output digital pulses are transmitted to a following TDC [6,7] for simultaneous digitization of both the charge and time information. Although this ASIC functions as expected [4,8], tests discovered significant inconsistencies among different chips, especially in the thresholds of the discriminators which is used for the time measurement. When integrating multiple channels onto a single chip, a high level of uniformity among the channels is indispensable. However, it was discovered that the thresholds of the discriminators in the ASIC could vary by as much as 10 times between the different chips, making it unacceptable for real applications. To address this issue, in this paper, we propose the application of an auto-compensation method to automatically adjust the working condition of the circuits inside the ASIC to achieve a high level of uniformity. In this work of research, a new version of the ASIC which integrates six channels within a single chip (two channels are combined together for the readout of one PMT) was designed and fabricated. Tests were conducted to evaluate the uniformity among the different channels and chips, as well as the overall charge and time measurements of the ASIC.

## 2. Discussion on the inconsistencies among multiple channels

To simplify the system structure, considering that both charge and time measurements are required, the TOT method is preferred. Fig.1 shows the architecture of a single channel of the front end readout ASIC. As shown in the figure, the input signals from PMTs are processed by the Charge-To-Time Converters (QTCs), to generate pulsed signals ("Charge Output" in Fig.1). The width of these pulsed signals correspond to the charge information of the input signals, while the signals time information are picked off through discriminators (Current Discriminator shown in Fig.1). These pulses could be digitized by the TDCs so that both charge and time measurements are achieved simultaneously [4,9,10]. Since nowadays TDCs can be easily designed based on Field-Programmable Gate Array (FPGA) devices [6,7], it will greatly simplify the system structure if all the QTC circuits are integrated within the ASICs.

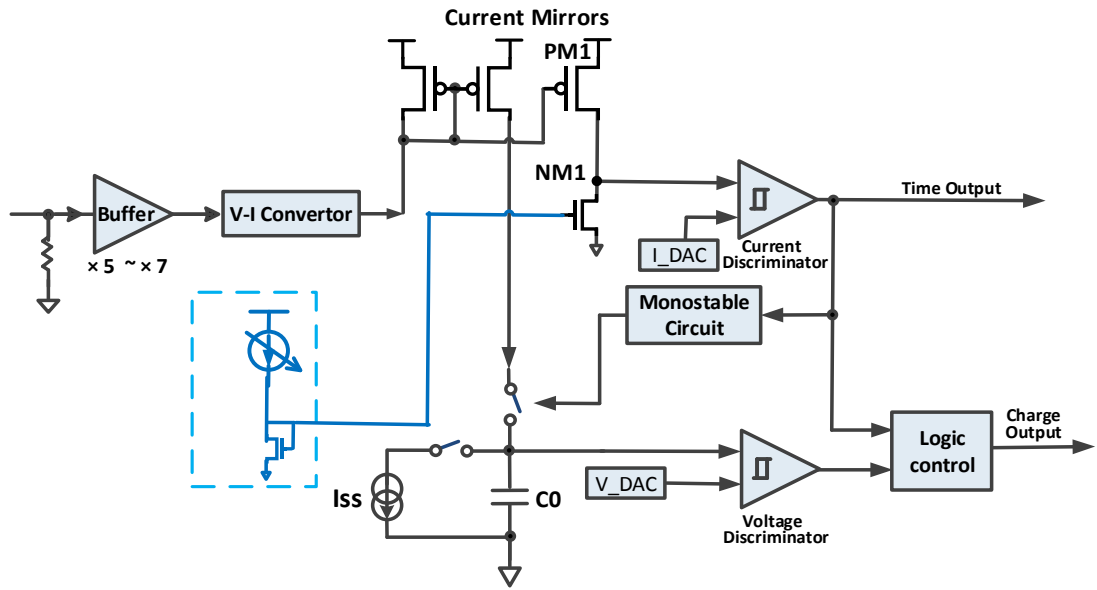

**Fig.1:** Achitecture of the WCDA front end readout ASIC (single channel).

To achieve a large dynamic range from 1 P.E. to 4000 P.E., two channels are employed to read out the signals from the anode and the dynode of one PMT, respectively. While the time measurement in the full-scale range is performed using the anode channel, the charge measurement is performed using a combination of the anode and dynode channels. For the charge measurement, the anode channel covers the amplitude range from 1 P.E. to 140 P.E. and the dynode channel covers the range from 30 P.E. to 4000 P.E. To guarantee a good noise performance and the measurement range of each channel, the current-mode signal processing method is used in ASIC design [4,5,11]. In the setup of the detector in WCDA, the output signals from the PMTs are transmitted to the ASIC using coaxial cables of 30 meters length. Since the input signal has a large dynamic range, a high precision impedance match is required which is achieved by using an external 50 Ω terminal resistor in our system. Meanwhile this 50 Ω resistor converts the current mode PMT signal to the voltage mode. The voltage mode signal is then amplified using a low-noise amplifier, and converted back to the current mode using a V-I converter for further processing. The output signal from the V-I converter in Fig.1 is duplicated over the two paths using a current mirror. One signal is fed into a current discriminator for performing the time measurement, while the other is converted to a digital pulse based on the linear discharging method. The width of this digital pulse is proportional to the input charge.

In the previous version prototype ASIC [4], we integrated two pairs of channels in a single chip. With the goal of optimizing the scheme by comparing the performances, each pair was designed using different methods for implementing the key circuits [4].

As mentioned above, test results indicate that while the previous prototype functions as expected, the inconsistencies among the channels can act as obstacles for real application. Especially, for the discriminators, whose thresholds must be set to equal to ¼ PE input signal for time measurement, the value of the thresholds might have a variance greater than 10 times among the channels.

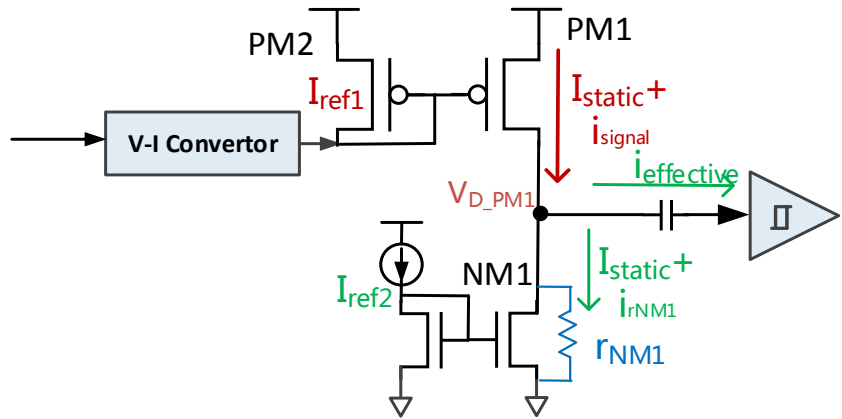

**Fig.2:** Detailed structure of the circuits between the V-I converter and the current discriminator.

To address this issue, this research devotes efforts to analyzing and optimizing the circuit structure. As shown in Fig.1, the signal is converted into the current mode using a V-I convertor, and then copied using a current mirror (shown as the PM1 and the NM1 branches). Fig.2 shows the details of the circuit structure between the V-I converter and the current discriminator. It shows that the output of the current mirror is actually the drain current of PM1 which is the sum of the static current ($I_{static}$) and the signal current ($i_{signal}$). The signal current ($i_{signal}$) is the dynamic current introduced by the input signal. The drain current of PM1 is split into two paths: one path feeds the NM1 (marked as "$I_{static}+i_{rNM1}$" in Fig.2), and the other path feeds the current discriminator (marked as "$i_{effective}$" in Fig.2). In order to achieve a high consistency among the channels, it is required that for the same input signal, the value of ieffective should be the same among all the channels. Actually, the drain voltage of PM1 (marked as "$V_{D\_PM1}$" in Fig.2) has a significant impact on $i_{effective}$ due to two factors. The first factor is that $V_{D\_PM1}$ is the voltage of the output of the current mirror, so $i_{signal}$ varies with different $V_{D\_PM1}$. The second factor is that the ratio of $i_{\text{effective}}$ and $i_{\text{rNM1}}$ depends on the ratio of the NM1 output resistor ($r_{\text{NM1}}$) and the equivalent input resistance of the current discriminator. $V_{\text{D\_PM1}}$ also has a direct influence on $r_{\text{NM1}}$. Therefore, $V_{\text{D\_M1}}$ also affects the ratio between the $i_{\text{effective}}$ and the $i_{\text{rNM1}}$.

In our previous design [4], as shown in Fig.2, a bias circuit was designed to feed a fixed external current $I_{\text{ref2}}$ into the system. In this scheme, however, $i_{\text{effective}}$ is strongly dependent on PVT (process, voltage and temperature) variation. This is due to the sensitivity of $V_{\text{D\_PM1}}$ to PVT variation. Actually we found that the PVT variation of difference between $I_{\text{ref2}}$ and $I_{\text{ref1}}$ in Fig.1 has crucial influence on the $V_{\text{D\_PM1}}$. This process can be analysed as follows.

When PM1 is in saturation mode, its drain current can be expressed as [12]

$$I_{static} = \frac{\mu_p C_{ox}}{2} \frac{W_{PM1}}{L_{PM1}} V_{ovPM1}^2 [1 + \lambda(V_{DSsatp} - V_{D\_PM1})] \quad (1),$$

where $\mu_p$ is the charge-carrier effective mobility, $C_{ox}$ is the oxide capacitance, $W_{PM1}$ and $L_{PM1}$ are the gate width and length of PM1, $V_{ovPM1}$ is the overdrive voltage of PM1, $V_{DSsatp}$ is the drain voltage of PM1 at the transition point from triode to saturation mode, and $\lambda$ is the channel-length modulation parameter.

As for PM2, its drain current $I_{\text{ref1}}$ can be expressed as

$$I_{ref1} = \frac{\mu_p C_{ox}}{2} \frac{W_{PM2}}{L_{MP2}} V_{ovPM2}^2 [1 + \lambda V_{pth}] \quad (2).$$

As PM1 and PM2 actually constitute a current mirror structure, the relationship between these two MOSFETs can be expressed as

$$\frac{W_{PM2}}{L_{PM2}} = \frac{1}{k_p} \frac{W_{PM1}}{L_{PM1}}, V_{ovPM2} = V_{ovPM1} \quad (3),$$

where $k_p$ is the gain of current mirror.

According to equation (1), (2), (3), the relationship between $I_{static}$, $I_{ref1}$, and $V_{D\_PM1}$ can be obtained as

$$I_{static} = I_{ref1} \frac{1 + \lambda(V_{DSsatp} - V_{D\_PM1})}{1 + \lambda V_{pth}} k_p \quad (4).$$

Analyzing NM1 in a similar way, we can further get

$$I_{static} = I_{ref2} \frac{1 + \lambda(-V_{DSsatn} + V_{D\_PM1})}{1 + \lambda V_{nth}} k_n \quad (5).$$

In GLOBALFOUNDRIES 0.35μm CMOS technology, $\lambda$ is around 0.1 ~ 0.01 V$^{-1}$, and $V_{th}$ is about 0.5-0.7 V. Thus $\lambda V_{th}$ in (5) is very small. Finally, $V_{D\_PM1}$ can expressed as

$$V_{D\_PM1} \approx \frac{I_{ref1} k_p - I_{ref2} k_n}{\lambda(I_{ref1} k_p + I_{ref2} k_n)} + V_{D\_PM1\_normal} \quad (6),$$

where $V_{D\_PM1\_normal}$ is the drain voltage of PM1 when $I_{ref1}k_p = I_{ref2}k_n$, i.e. the ideal value of $V_{D\_PM1}$ in our previous design.

In our design $k_p$ equals to $k_n$, which means

$$V_{D\_PM1} \approx \frac{I_{ref1} - I_{ref2}}{\lambda(I_{ref1} + I_{ref2})} + V_{D\_PM1\_normal} \quad (7).$$

As mentioned above, λ is around 0.1 ~ 0.01 V$^{-1}$, while ($I_{ref1}$+ $I_{ref2}$) is about 100 μA in our desgin, so the total gain from ΔI ($I_{ref1}$ - $I_{ref2}$) to $V_{D\_PM1}$ is around 100~1000 kΩ. Therefore, even a tiny difference between $I_{ref1}$ and $I_{ref2}$ due to PVT variation will lead to a large dispersion of $V_{D\_PM1}$, which is the cause of the inconsistencies among channels.

These relationships between the components have been confirmed by tests conducted on the previous version of the ASIC. Since $I_{ref1}$ is fixed by the ASIC circuits to around 50 μA, we made modifications to the value of $I_{ref2}$ to observe the relationship between the discriminator threshold (corresponding to $i_{effective}$) and ΔI ($I_{ref2}$- $I_{ref1}$).

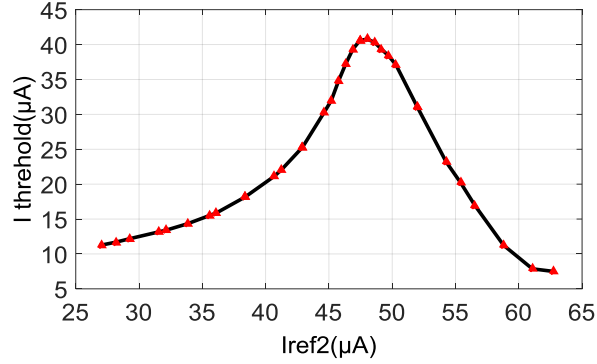

**Fig.3**: Test results of the relationship between threshold and $I_{ref2}$.

As shown in Fig.3, test results indicate that when $I_{ref2}$ is changed from 25 μA to 65 μA, the threshold varies from 6 μA to 42 μA. We also tested the drain voltage of PM1 with different $I_{ref2}$. In Fig.4 we can see the slope of the curve is about 100kΩ when $I_{ref2}$ is set to around the ideal value of 50 μA, which agrees with the expression derived in equation (7).

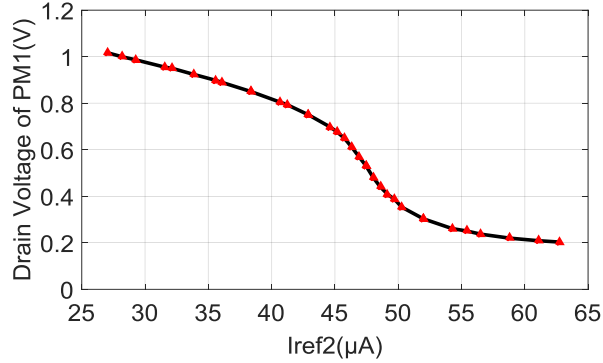

**Fig.4:** Test results of the relationship between $V_{D\_PM1}$ and $I_{ref2}$.

In fact, due to PVT variations, there exist quite considerable variations in $I_{ref1}$ among different channels. This is because, as shown in Fig.1, in our previous version of the ASIC, a low noise buffer with high gain was used to guarantee a good noise performance in the circuit. The input offset of the buffer, which is introduced by the variation in the PVT, is amplified by itself and finally leads to a large random output offset. This voltage offset is further converted to the current offset of $I_{ref1}$ through a 750 Ω resistor in the V-I converter [4] in Fig.1. In Fig.5, we can see the results of the Monte Carlo simulations. These simulations reveal obvious variations in the offset of the buffer output, which can cause considerable variations in $I_{ref1}$.

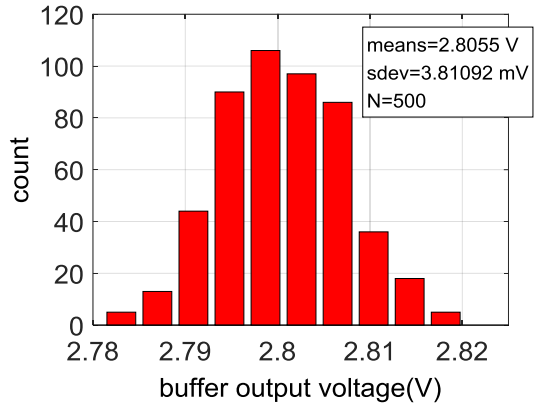
**Fig.5:** Histogram simulation results of the buffer output voltage.

Due to the effect of these variations, the curve in Fig.3 will disperse into multiple curves in the actual channels of the ASIC. The results of further simulations conducted to evaluate the influence of these variations are plotted in Fig.6. In the figure, we can see the relationship between the threshold and the $I_{ref2}$. The different curves correspond to the different values of $I_{ref1}$ caused by the PVT variation, and we can observe a significant dispersion. This indicates, that if $I_{ref2}$ is set to the same value for all channels, we would observe obvious inconsistencies in the thresholds.

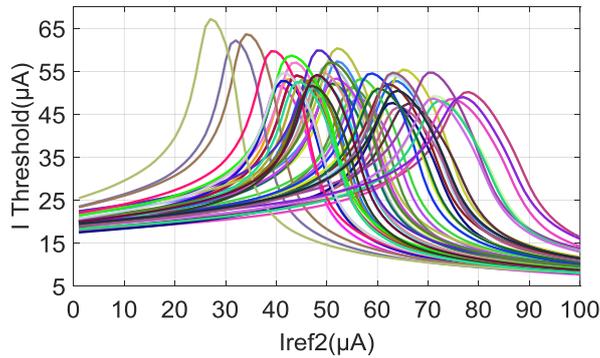
**Fig.6:** Monte Carlo simulation results of the relationship between the thresholds and $I_{ref2}$.

To address this issue, we found an alternative way and modified our circuit design. Since the value of ieffective depends on $V_{D\_PM1}$, we could consider fixing the value of $V_{D\_PM1}$ to obtain a better consistency among the channels. We conducted simulations to evaluate the performance of this method. As shown in Fig.7, based on these simulations we obtain a better dispersion among the curves compared with Fig.6. A higher consistency of the thresholds can be obtained by selecting an appropriate value of $V_{D\_PM1}$.

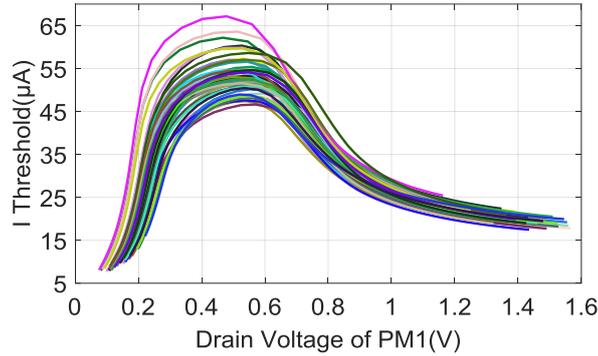

**Fig.7:** Monte Carlo simulation results of the relationship between threshold and $V_{D\_PM1}$.

## 3. Circuit design

### 3.1 Architecture of the ASIC with the auto-compensation function

According to the above analysis, a high consistency among the channels can be achieved through monitoring and controlling the voltage of the key point in the circuits ($V_{D\_PM1}$). Based on this idea, the circuits were designed to observe and tune $V_{D\_PM1}$ through automatic compensation. One advantage is that, this automatic compensation needs to be applied only for the first time the chip is used. In total, there are three pairs of channels in a single chip. To reduce the resource and power consumption of the chip, a shared bus architecture, as shown in Fig.8 was designed.

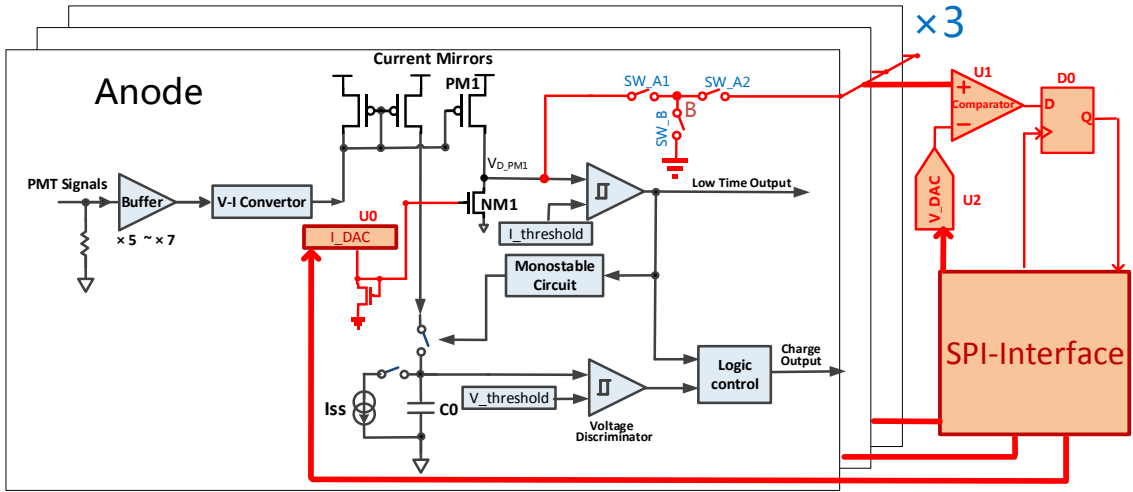

**Fig.8:** Auto-compensation circuit strucuture.

Three switches are added (SW_A1, SW_A2, and SW_B in Fig.8) to each channel of the ASIC. The compensation process begins when the ASIC is used for the first time. For the channel where we want to apply the compensation, the switches SW_A1 and SW_A2 are turned on and the switch SW_B is turned off. We apply a reverse configuration (turn off SW_A1, SW_A2 and turn on SW_B) to the other channels. Under this configuration, the $V_{D\_PM1}$ of interested channel is connected to the shared bus, and can be further processed. The switches in each of the channels are alternately turned on to compensate for all the channels. Once the compensation process is completed, the switches SW_A1 and SW_A2 are turned off and the SW_B switch is turned on in all the channels. This three-switch structure is effective in avoiding crosstalk among the different channels. Shielding to other channels is built up by first grounding the point B using SW_B and then placing the other two cascaded switches (SW_A1 and SW_A2) at high resistance.

Other components of the compensation circuits include a comparator (U1 in Fig.8), a voltage DAC (U2 in Fig.8), a D-flip-flop (D0 in Fig.8), and a current DAC (U0 in Fig.8). $V_{D\_PM1}$ within the channel under compensation is fed to the comparator U1, and the threshold of U1 is the output of the DAC U2. Voltage DAC (U2) is set to a proper value by configuring it via the Serial Peripheral Interface (SPI) interface through an external FPGA. This SPI interface can also be used to read the flip-flop D0 which stores the output of the comparator (U1). Then the current DAC (U0) is used to tune the drain current of NM1 until $V_{D\_PM1}$ is approximate to the predetermined voltage (the output of the voltage DAC (U2)). Since $V_{D\_PM1}$ is the same for each channel, this allows to achieve a high consistency among the channels. The tuning process of the NM1 drain current is based the Bisection method. The corresponding logic design is illustrated in Fig.9.

This circuit can also be used to quantize the value of $V_{D\_PM1}$ in debug mode by tuning the voltage DAC (U2) instead of the current DAC (U0).

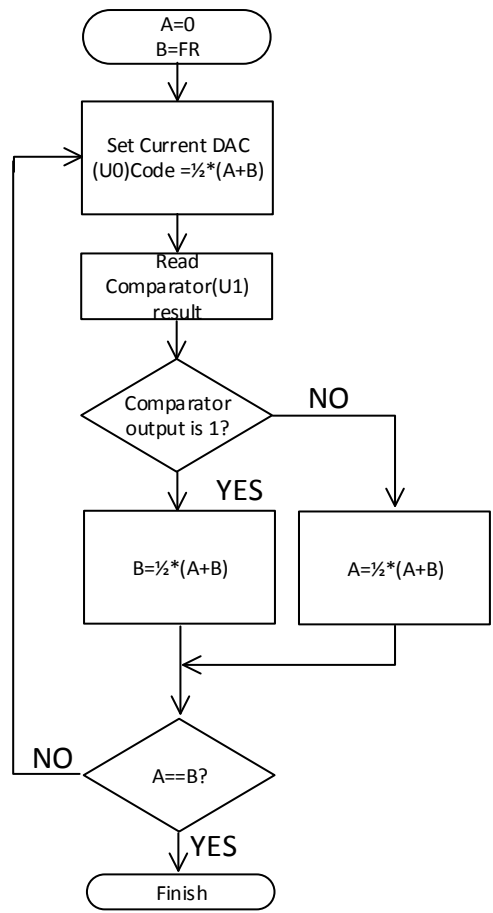

**Fig.9:** Logic design for the auto-compensation structure. A and B are endpoint values in the Bisection method. FR is the full range of the current DAC.

In order to further decrease power consumption, the compensation circuit inside the ASIC can be shut down after the compensation process is over.

A few special considerations must be taken while designing the current DAC. Since the drain current of NM1 is tuned through this DAC, in order for the procedure in Fig. 9 to work correctly, the transfer curve of the current DAC is required to be monotonous. This means its Differential Non-Linearity (DNL) must be smaller than 1 LSB.

**3.2 Non-Binary Weighted current DAC**

In order to ensure monotonicity, a non-binary weighted current DAC is implemented in the chip. Firstly, we considered a normal binary weighted DAC structure as shown in Fig.10 (a). The DNL of this kind of DAC is introduced by a mismatch variation in the fabrication process. A Monte Carlo simulation was conducted to estimate this effect. As shown in Fig.12 (a), the worst DNL of this binary weighted DAC appears in code 6'b100000 (half of full range) with a value of 1.28 LSB (>1 LSB), which is unacceptable in this case.

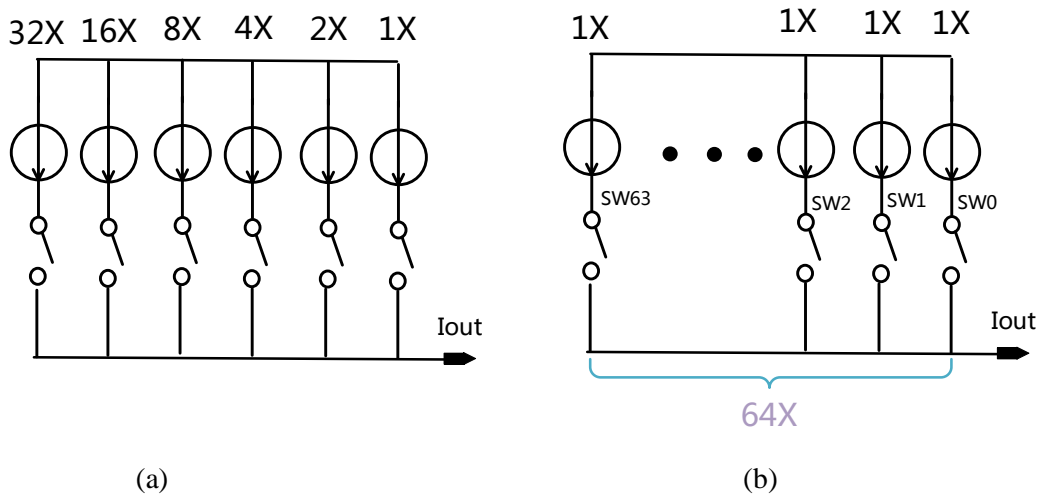

**Fig.10:** Two structures of current DAC-(a) Binary weighted current DAC; (b) Unit weighted DAC.

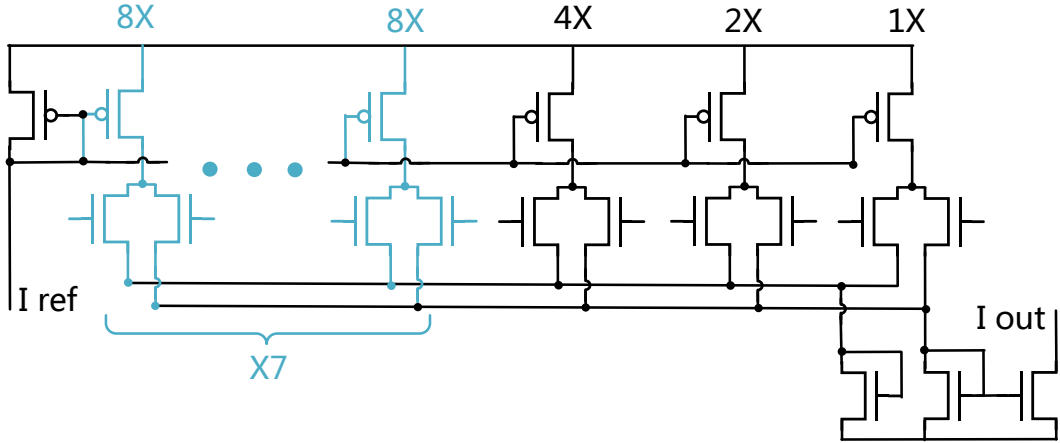

**Fig.11:** Structure of the combined weighted DAC.

As shown in Fig.10 (b), the unit weighted DAC, which has only a single changing switch in each transition, can obtain a uniform DNL. This DNL will theoretically be smaller than that of a normal binary weighted DAC. However, since a unit weighted DAC requires more switches (up to 64 switches in our application), and a complicated decode circuit to translate the binary code to the temperature code. These will increase complexity of the circuit.

Therefore, a combined structure is preferred in this case [13]. Fig.12 (a) shows that the DNL exceeds 1 LSB only in three transition codes (100000, 010000, and 110000). Therefore, in consideration of the above, as shown in Fig 11, we decided to replace the 3 most significant bits (MSB) of the binary weighted DAC with the unit weighted structure, and retained the binary weighted structure for the lowest 3 bits. Using this combination, we were able to achieve a high DNL performance while maintaining circuit simplicity.

As shown in Fig.12 (b), the value of the DNL for all codes is now less than 0.7 LSB, and only 10 switches and a 3-bit decode circuit are needed.

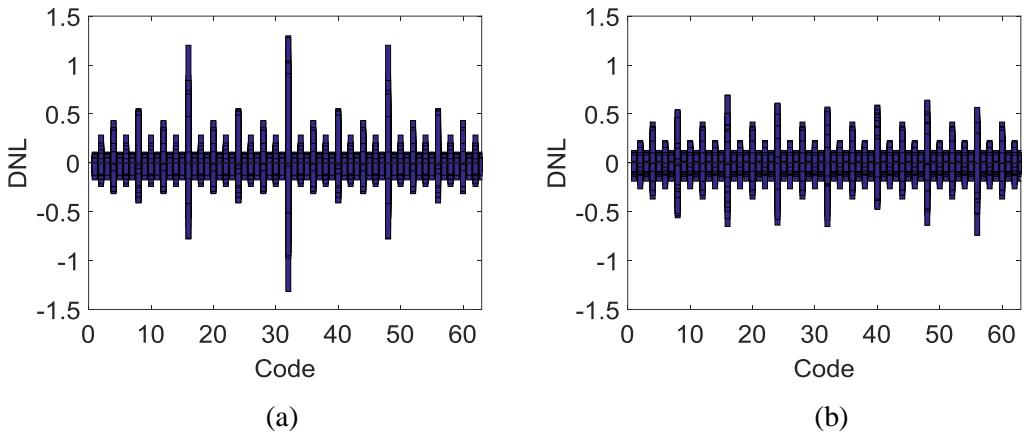

(a)                  (b)

**Fig.12:** Monte Carlo simulation results of DNL-(a) DNL of the binary weighted current DAC; (b) DNL of the combined weighted current DAC.

**3.3 Application-specific SPI interface for the multichannel chip**

Since using our proposed architecture, the entire chip only needs to be compensated once before use, only a flexible configuration and monitor interface needs to be integrated inside the ASIC, and most of the logical parts can be placed off the chip to save area and power.

In the revised version of the ASIC, there are a total of 6 signal channels and an auto-compensation circuit. A customized SPI based data interface is implemented to configure and monitor the 6 signal channels individually. This customized interface can be used in the auto-compensation process as well.

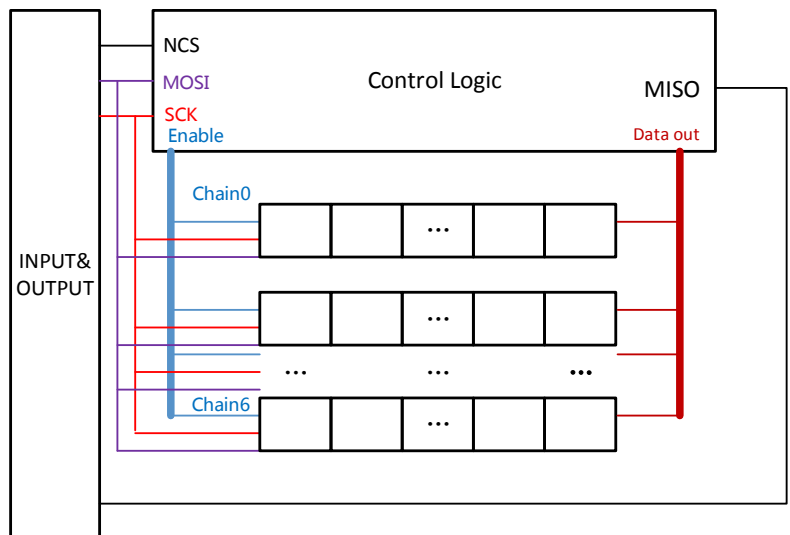

**Fig.13:** Struture of the SPI based interface.

As shown in Fig.13, the three main parts of the SPI interface are a) the input and output buffers, b) the control logic, and c) the register chains. There are 7 register chains in the chip, including 6 chains (Chain Nos. 0 to 5) used for the configuration of the 6 channels, and one chain (Chain No. 6) used for the bias and the auto-compensation circuits.

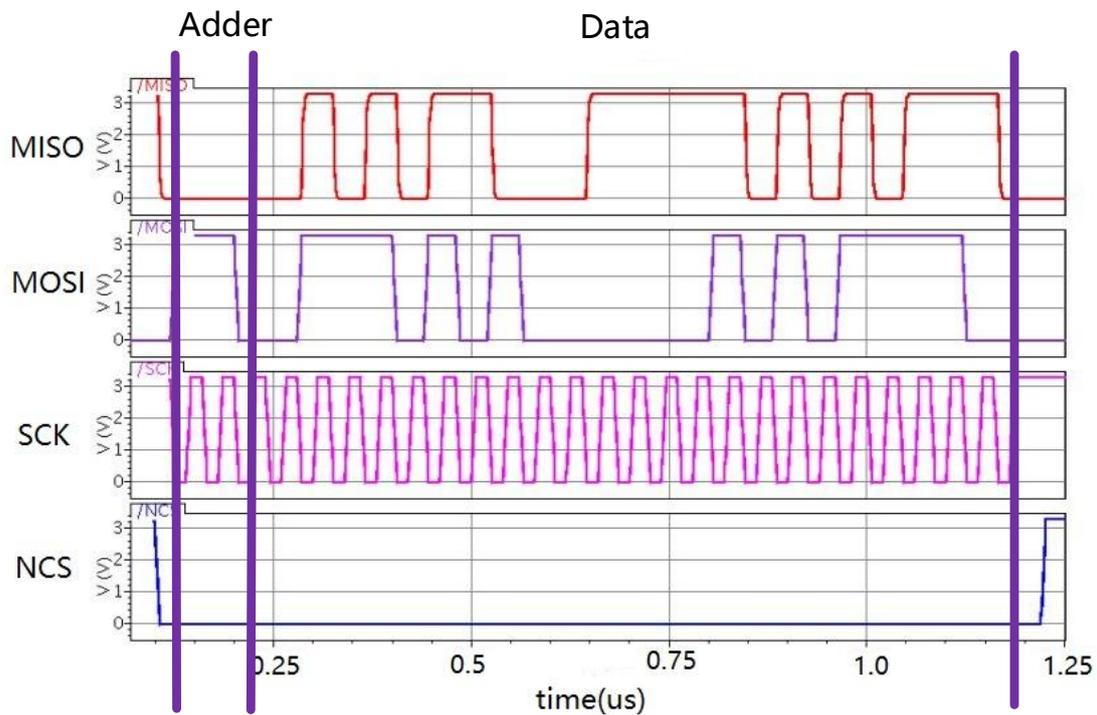

**Fig.14**: Simulation results of the SPI based interface timing diagram.

The simulation results of the SPI based interface timing diagram is shown in Fig.14. In our application, we use a standard SPI port with the default operational configuration [14]. The NCS signal is used to select and enable the chip. The SCK is the clock signal sent from the master to the ASIC. The MOSI signal is the data signal that the master sends to the ASIC, and the MISO signal is the data signal that the ASIC sends back to the master.

Cascaded registers are placed to form a chain in each channel and an address decoder is implemented in the data interface to maintain the individual configuration of each channel. The ASIC uses a custom configuration format where every command starts with a 3-bit address for choosing the corresponding register chain, and is followed by 24 bits used for storing the content.

## 4. Test Results

Based on the above analysis and circuit optimization scheme, a multiple channel ASIC was designed and fabricated in the GLOBALFOUNDRIES 0.35 μm CMOS technology. Fig.15 shows the photo of this new ASIC, which consists of 6 channels and an auto-compensation circuit. Tests were conducted to evaluate the performance of this ASIC, including the mismatch among the channels, as well as charge and time measurements.

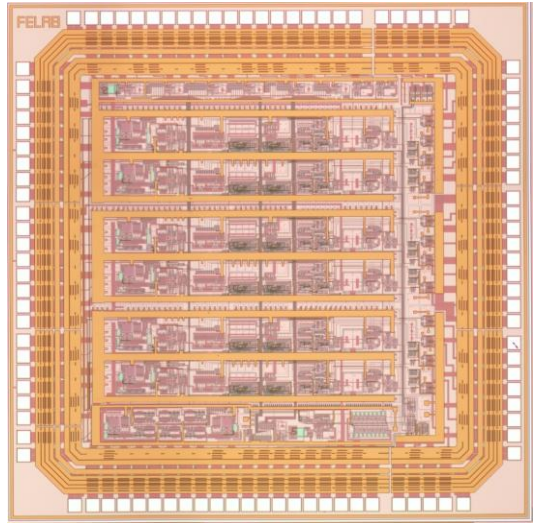

**Fig.15:** Photograph of the new ASIC.

A signal generator (Tektronix AFG3252) was used to generate the input signal for the ASIC according to the output waveform of the PMT (R5912, signal rise time ~ 8 ns).

### 4.1 Thresholds Mismatch Test Results

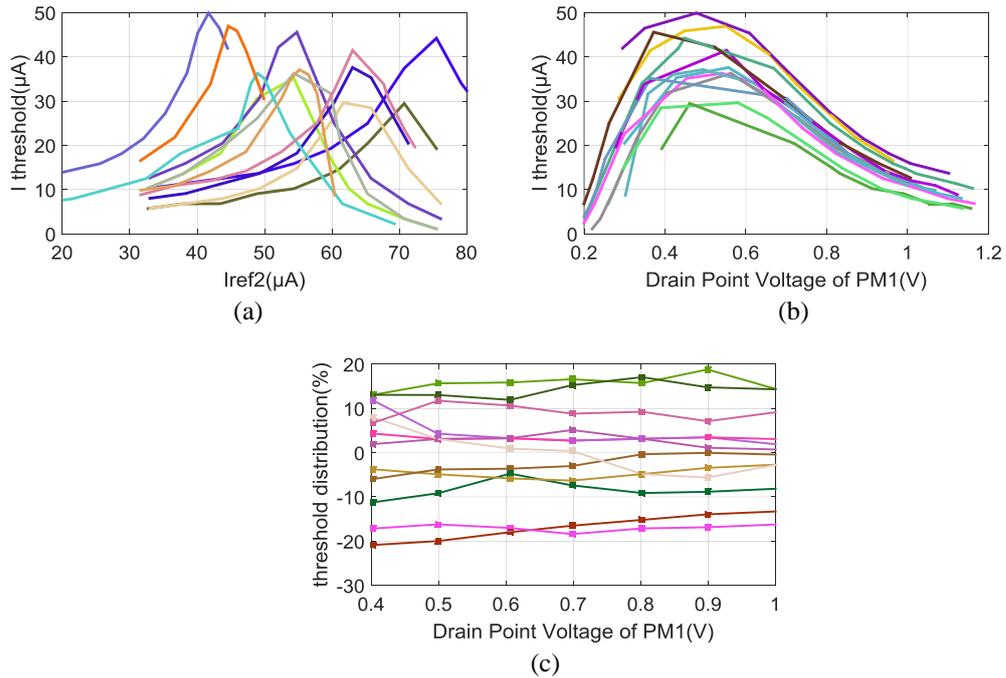

**Fig.16:** Thresholds mismatch test results-(a) relationship between threshold and $I_{ref2}$; (b) relationship between threshold and $V_{D\_PM1}$; (c) variation of threshold with different $V_{D\_PM1}$.

The S-curve method is used to obtain the threshold of each channel [15]. The Fig.16 (a) shows the change in the thresholds versus $I_{ref2}$ (in Fig.2). The evaluation are conducted for the 12 anode channels in 4 chips. We can observe that the scattering range of these curves is quite large. This observation agrees well with our analysis and our simulation results presented previously in Fig.6. Fig.16 (b) plots the thresholds of the channels against the VD_PM1, which also agrees well with our previous simulations (as shown in Fig.7). Fig.16 (c) shows the final threshold mismatch among the channels, and the result show that this mismatch is lower than 20%. This results for the improved version of the ASIC are much better than the results of the previous version (variation larger than 10 times), and satisfies the application requirements.

**4.2 Multichannel Charge Measurement Performance**

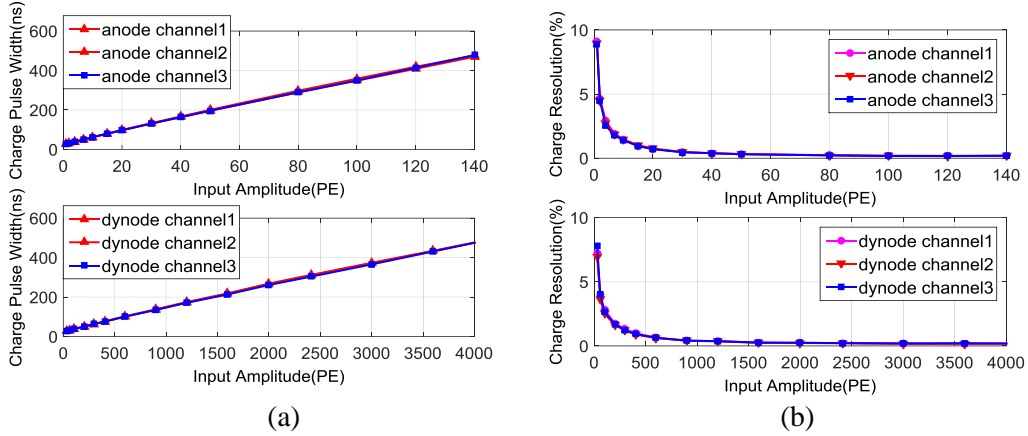

**Fig. 17:** Charge measurement performance test results-(a) charge measurement transfer curve; (b) charge measurement resolution.

The charge output of the ASIC (shown in Fig.1) is a digital pulse, whose width corresponds to the input charge information. As shown in Fig. 17(a), the anode and the dynode channels cover a dynamic range from 1 P.E. ~ 140 P.E. and 30 P.E. ~ 4000 P.E., respectively. Combining these two channel, a total dynamic range of 1 P.E. ~ 4000 P.E. with sufficient overlap is achieved. As shown in Fig.17 (b) shows the charge measurement resolution is better than 10% at 1 P.E. and 1% at 4000 P.E.

**4.3 Multichannel Timing Measurement Performance**

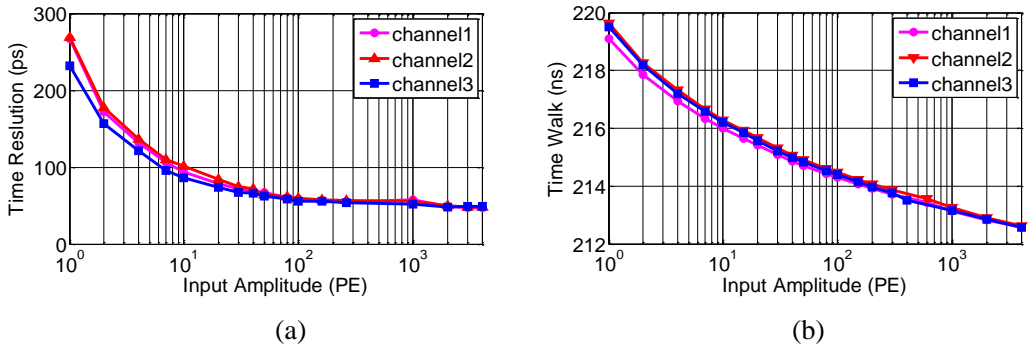

**Fig.18:** Time measurement performance test results-(a) time measurement resolution; (b) time walk.

To evaluate the time measurement performance of the ASIC, a Tektronix AFG3252 was used to generate two high precision synchronous signals, with an inter-signal jitter lower than 50 ps. One of the signals is transmitted to the ASIC using a 30m cable while the other is transmitted to an oscilloscope (Lecroy 104 MXi) as reference for the time measurement of the ASIC output. As shown in Fig.18, the time measurement resolution is better than 300 ps over a dynamic range of 1 P.E. ~4000 P.E., and the time walk is less than 8 ns.

## 5. Conclusion

Aiming at high precision charge and time measurement for WCDA in LHAASO, a multichannel current mode ASIC was designed. To address the mismatch among channels, a new auto-compensation structure is proposed to guarantee a good consistency for different channels. The results of the tests conducted to evaluate the performance of the ASIC indicate that the mismatch among the 12 channels of 4 chips is below 20%. The tests show that the charge measurement resolution is better than 10% at 1 P.E. and 1% at 4000 P.E., while the time measurement resolution is better than 300 ps. These performances have met the application requirements.

**Acknowledgments**

The authors would like to thank Allan K. Lan of the MD Anderson Cancer Center in University of Texas for his help regarding this work over the years. This work is supported by National Natural Science Foundation of China (11722545) and the CAS Center for Excellence in Particle Physics (CCEPP).

**References**

[1] LHAASO collaboration, Z. Cao, *A future project at Tibet: The large high altitude air shower observatory (LHAASO)*, Chin. Phys. **C 34** (2010) 249.

[2] M.J. Chen et al., *R&D of LHAASO-WCDA*, in *Proceedings of 32nd International Cosmic Ray Conference*, Beijing, China, 2011.

[3] L. Zhao, S.-B. Liu and Q. An, *Proposal of the readout electronics for the WCDA in the LHAASO experiment*, Chin. Phys. **C 38** (2014) 016101.

[4] L. Zhao et al, *Prototype of a front-end readout ASIC designed for the Water Cherenkov Detector Array in LHASSO, JINST.* 10 (2015) no.03, P03015

[5] Deng Z, Lan A K, Sun X, et al., *Development of an eight-channel time-based readout ASIC for PET Applications,* IEEE Transactions on Nuclear Science, 2011, 58(6): 3212-3218.

[6] J. Wu and Z. Shi, *The 10-ps wave union TDC: improving FPGA TDC resolution beyond its cell delay,* IEEE Nucl. Sci. Symp. Conf. (2008) 3440.

[7] L. Zhao et al., *The design of a 16-channel 15 ps TDC implemented in a 65 nm FPGA,* IEEE Trans. Nucl. Sci. 60 (2013) 3532

[8] Wei-Hao Wu et al., *Evaluation of a front-end ASIC for the readout of PMTs over a large dynamic range,* Chin.Phys. C39 (2015) no.12, 126101

[9] H. Nishino, K.Awai, and Y.Hayato et. al., *High-speed charge-to-time converter ASIC for the Super-Kamiokande detector*, Nucl. Instrum. Meth. A 610 (2009)710.

[10] G. Martin-Chassard et al., *PARISROC, a photomultiplier array readout chip (PMm2 collaboration)*, Nuclear Instruments and Methods in Physics Research A 623 (2010) 492–494.

[11] X. Zhu et al., *TIMPIC-II: the second version time-based-readout ASIC for SSPM based PET applications*, IEEE Nucl. Sci. Symp. Conf. Rec. (2012) 1474.

[12] Behzad Razavi, *Design of Analog CMOS Integrated Circuits,* MC Graw Hill Education ISBN-10: 0072380322

[13] Hyuen-Hee Bae et al., *A 3 V 12b 100 MS/s CMOS D/A CONVERTER FOR HIGH-SPEED SYSTEM APPLICATIONS* Circuits and Systems, 2003. ISCAS '03. Proceedings of the 2003 International Symposium on, 2003, pp. I-869-I-872 vol.1.

[14] F. Leens, *An introduction to I2 C and SPI protocols* IEEE Instrumentation & Measurement Magazine, vol.12, no.1, pp.8-13, February 2009.

[15] A. Comerma et al., *Front End ASIC design for SiPM readout* Journal of Instrumentation  2016  11 P10005